\documentclass{article}
\usepackage{spconf,amsmath,amssymb,amsfonts,bm,mathtools,graphicx}
\usepackage{algorithm,algorithmic,cite}
\usepackage{url}

\graphicspath{{../figures/}}

\def\Dmax{D_{\mathrm{max}}}
\def\dmove{\Delta_{\max}}
\def\rmin{r_{\mathrm{min}}}
\def\Fmat{\mathbf{F}_{\mathcal{M}}}
\def\Sig{\mathbf{\Sigma}}

\title{TDOA-Based Online Target Tracking with \\ Simultaneous Sensor Pairing and Relocation}

\name{Ryosuke Ikura$^{1}$, Junya Hara$^{1}$, Hiroshi Higashi$^{2}$, Yuichi Tanaka$^{1}$
\thanks{This work is supported in part by JSPS KAKENHI under Grant 26H02536, and JST AdCORP under Grant JPMJKB2307.}}
\address{$^{1}$The University of Osaka, Osaka, Japan~~
  $^{2}$Kansai University, Osaka, Japan}

\begin{document}
\ninept
\setlength{\textfloatsep}{10pt plus 2pt minus 2pt}
\setlength{\dbltextfloatsep}{10pt plus 2pt minus 2pt}
\setlength{\abovedisplayskip}{4pt plus 1pt minus 1pt}
\setlength{\belowdisplayskip}{4pt plus 1pt minus 1pt}
\setlength{\abovedisplayshortskip}{2pt plus 1pt}
\setlength{\belowdisplayshortskip}{2pt plus 1pt}
\renewcommand{\dbltopfraction}{0.92}
\renewcommand{\dblfloatpagefraction}{0.7}
\renewcommand{\topfraction}{0.92}
\renewcommand{\floatpagefraction}{0.7}
\renewcommand{\textfraction}{0.06}
\maketitle

\begin{abstract}
We propose a target tracking method for mobile sensor networks (MSNs) based on time difference-of-arrival (TDOA).
Target tracking using an MSN requires reorganizing sensor pairs and relocating sensor positions to obtain TDOAs with higher quality.
Existing studies only consider either one while fixing the other, which may limit the tracking accuracy.
We address this limitation by simultaneously solving sensor pairing and relocation.
We formulate the problem as a maximization of the determinant of the Fisher information matrix, and decompose the problem into two subproblems to solve them alternately.
Sensor pairing is solved by a mixed-integer second-order cone program algorithm, and sensor relocation is solved by majorization-minimization.
Experimental results show that the proposed method obtains lower tracking error than existing designs with a practical computation time.
\end{abstract}


\section{Introduction}
\label{sec:intro}
Mobile sensor networks (MSNs) play a vital role in modern IoT applications~\cite{shorey2006mobile,munir2007mobile}.
Sensor networks consist of sensors and communication links between them~\cite{puccinelli2005wireless,sohraby2007wireless}.
In mobile cases, sensor positions and links among them can be time-varying.

Target tracking is one of the most important applications of MSNs.
Its aim is to localize and track the position of moving target(s) from sensor measurements~\cite{meng2016optimal,meng2013decentralized}.
In practice, targets are often noncooperative and sensors and targets may work in an unsynchronized manner~\cite{ho1993solution}.
Therefore, their localization should rely on measurements invariant to transmit-time shifts~\cite{so2011source,li2016contributed}.
Time-difference-of-arrival (TDOA) is one of such possible measurement domains.
It is defined as the difference between the arrival times of a signal at two sensors~\cite{ho1993solution}.
Since measuring TDOAs on all sensor pairs could be impractical in terms of energy consumption and communication overhead~\cite{meng2013decentralized,martin2010algorithms}, only a subset of the pairs is activated~\cite{yaqin2025sensor}.

TDOAs are often degraded by noise mainly from two sources.
1) Hardware limitation of the sensors, such as clock jitter.
It typically produces additive white Gaussian noise~(AWGN)~\cite{li2007jitter}.
2) Geometry-induced noise stems from the observation environment, such as multipath propagation.
It is modeled as AWGN with variance proportional to the target-to-sensor distances~\cite{dardari2009ranging,zhao2019sensor}.
As a result, TDOA-based tracking has to address the combination of these two noises.

TDOA-based tracking is performed through two phases: Measurement acquisition and target localization~\cite{martalo2023hybrid,kim2019efficient}.
In the first phase, the sensors are relocated from current positions, and/or sensor connections of sensors are reorganized, i.e., sensor pairs.
In the second phase, the target is localized from the measured TDOAs by using a maximum likelihood estimator~\cite{mahfouz2014target,ma2021maximum,foy1976position,torrieri2007statistical}.
Note that the tracking accuracy directly depends on the sensor pairs and the sensor positions.

Sensor pairing and sensor relocation are studied in different lines of research~\cite{yaqin2025sensor,khalil2026resourceawaretopologymanagementisacenabled,meng2016optimal,meng2013decentralized}.
An existing sensor pairing method~\cite{yaqin2025sensor} selects pairs while keeping the sensor positions fixed.
When the targets are far from sensors, measurements may be severely degraded by geometry-induced noise.
In contrast, an existing sensor relocation method~\cite{meng2016optimal} updates the position of each sensor while pairs are fixed.
With such fixed pairs, the tracking accuracy can be improved only by moving the sensors under movement constraints.
These motivate us to consider simultaneously optimizing sensor pairing and relocation for accuracy improvement of target tracking.

In this paper, we propose a simultaneous pairing and relocation method of MSNs based on TDOA.
We formulate the sensor pairing and the sensor relocations as one problem that maximizes the D-optimality criterion of the Fisher information matrix~(FIM) evaluated at the current target position.
We then alternately solve each subproblem while keeping the other fixed.
The pairing is cast as a mixed-integer second-order cone program~(MISOCP)~\cite{ikura2026misocp}.
The relocation is solved by majorization-minimization under constraints on the sensor movement.

In simulations of target tracking, the proposed method obtains lower time-averaged root mean squared error~(RMSE) than existing methods that optimize only the pairing or only the relocation.
We also demonstrate that the proposed method runs in a practical time for online tracking.

\textit{Notation:} Bold lowercase and uppercase letters denote vectors and matrices, respectively.
$[\mathbf{X}]_{ij}$ is the $(i,j)$ element of $\mathbf{X}$, $[\mathbf{X}]_{a}$ its $a$th column, and $[\bm{x}]_i$ is the $i$th element of $\bm{x}$.
Calligraphic letters represent sets, and $^{\top}$ denotes the transpose.
$\|\cdot\|$ is the $\ell_2$ norm, $\nabla$ the gradient, $\bm{1}$ the all-ones vector, $\operatorname{diag}(\cdot)$ the diagonal matrix carrying the elements of its vector argument, and $\Pi_{\mathcal{C}}(\cdot)$ the Euclidean projection onto a set $\mathcal{C}$.

\begin{figure*}[t]
  \centering
  \includegraphics[width=0.85\textwidth]{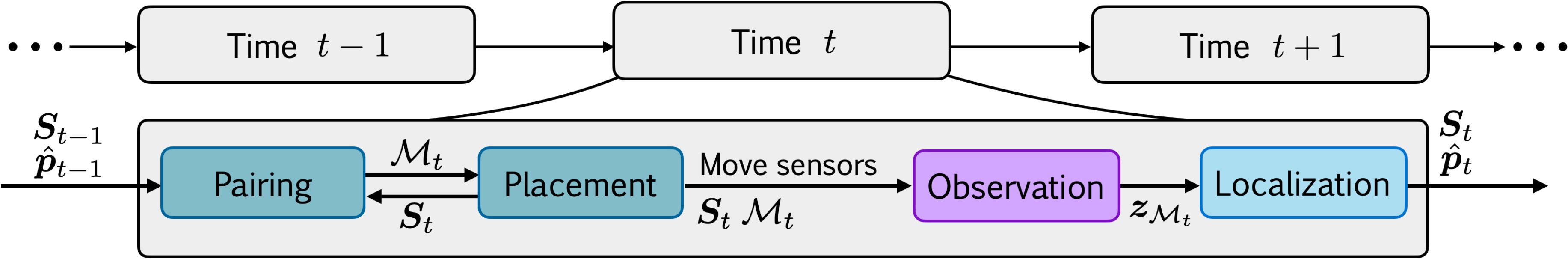}
  \caption{Online target tracking framework.
  The bottom row details the operations at time $t$.}
  \label{fig:framework}
\end{figure*}

\section{Preliminaries}
\label{sec:prelim}
We consider a two-dimensional setting with a single target at the true position $\bm{p}\in\mathbb{R}^2$, and we write the $i$th sensor position as $\bm{s}_i\in\mathbb{R}^2~(i=1,\ldots,N)$.
The $N$ sensor positions are stacked into $\bm{S}=[\bm{s}_1,\dots,\bm{s}_N]^\top$.
For pair $k$ that connects sensors $i$ and $j$, the noiseless TDOA is
\begin{equation}
\label{eq:tau}
  \tau_{k}(\bm{p})=(\|\bm{p}-\bm{s}_i\|-\|\bm{p}-\bm{s}_j\|)/c,
\end{equation}
where $c$ is the propagation speed~\cite{yaqin2025sensor}.
Without loss of generality, we set $c=1$.

In this paper, we assume that TDOA is observed with AWGN as $z_{k}(\bm{p})=\tau_{k}(\bm{p})+n_{k}$, where $n_{k}\sim\mathcal{N}\big(0,\sigma_{k}^2(\bm{p})\big)$ is independent across pairs.
In practice, TDOA measurements are degraded by two types of error: Gaussian noise with a fixed variance due to clock jitter~\cite{li2007jitter,dardari2009ranging}, and one with a noise increases with the target-to-sensor distance~\cite{zhao2019sensor,huang2015tdoa}.
Therefore, we model the variance $\sigma_{k}^2(\bm{p})$ as
\begin{equation}
\label{eq:noise}
  \sigma_{k}^2(\bm{p})=\sigma_{\min}^2+\sigma_0^2\big(\|\bm{p}-\bm{s}_i\|^2+\|\bm{p}-\bm{s}_j\|^2\big),
\end{equation}
where $\sigma_{\min}^2$ corresponds to noise for clock jitter and $\sigma_0^2$ scales the term that grows with the target-to-sensor distances~\cite{zhao2019sensor}.

Let $\mathcal{M}$ be the set of active pairs, and stack the TDOAs over $\mathcal{M}$ into $\bm{\tau}_{\mathcal{M}}(\bm{p})=[\tau_k(\bm{p})]_{k\in\mathcal{M}}$ and $\bm{z}_{\mathcal{M}}(\bm{p})=[z_k(\bm{p})]_{k\in\mathcal{M}}$.
Since the noise is independent across pairs, the covariance of $\bm{z}_{\mathcal{M}}(\bm{p})$ is $\Sig_{\mathcal{M}}(\bm{p})=\operatorname{diag}\big(\{\sigma_k^2\}_{k\in\mathcal{M}}\big)$.
The measurements are modeled as $\bm{z}_{\mathcal{M}}(\bm{p})\sim\mathcal{N}\big(\bm{\tau}_{\mathcal{M}}(\bm{p}),\Sig_{\mathcal{M}}(\bm{p})\big)$.

The accuracy of estimating $\bm{p}$ from $\bm{z}_{\mathcal{M}}(\bm{p})$ is characterized by the FIM of $\bm{p}$~\cite{vantrees2001detection}.
Since both the mean and the covariance depend on $\bm{p}$, the FIM is given by the Slepian--Bangs formula~\cite{kay1993statistical,torrieri2007statistical}.
Let $\mathbf{J}_{\mathcal{M}}$ be the Jacobian of $\bm{\tau}_{\mathcal{M}}(\bm{p})$ with respect to $\bm{p}$.
The $k$th row of $\mathbf{J}_{\mathcal{M}}$ is written as $\bm{w}_k^\top\triangleq(\bm{u}_i-\bm{u}_j)^\top$, where $\bm{u}_i$ and $\bm{u}_j$ are the unit bearing vectors from each sensor to the target, and $\bm{u}_i=(\bm{p}-\bm{s}_i)/\|\bm{p}-\bm{s}_i\|$.
The $(a,b)$ entry of the $2\times 2$ matrix $\mathbf{F}_{\mathcal{M}}(\bm{p},\bm{S})$ is given by
\begin{equation}
\label{eq:fimT}
  [\mathbf{F}_{\mathcal{M}}(\bm{p},\bm{S})]_{ab}
  =[\mathbf{J}_{\mathcal{M}}]_{a}^{\top}\Sig_{\mathcal{M}}^{-1}[\mathbf{J}_{\mathcal{M}}]_{b}
  +\tfrac{1}{2}\operatorname{tr}\!\big(\Sig_{\mathcal{M}}^{-1}\Sig_{\mathcal{M},a}
  \Sig_{\mathcal{M}}^{-1}\Sig_{\mathcal{M},b}\big),
\end{equation}
where $\Sig_{\mathcal{M},a}=\partial\Sig_{\mathcal{M}}/\partial[\bm{p}]_a$, and every quantity on the right-hand side is evaluated at $\bm{p}$ and $\bm{S}$.
The Cram\'er--Rao bound~(CRB) at the true position is $\Fmat^{-1}(\bm{p},\bm{S})$~\cite{vantrees2001detection}, and hence maximizing an optimality criterion of FIM is equivalent to minimizing the lower bound on the variance of an unbiased estimator of $\bm{p}$.

\section{Related Work}
\label{sec:related}
This section summarizes the existing designs of sensor pairing and sensor relocation.

\noindent
\textbf{Sensor Pairing}: 
Sensor pairing methods select a subset of sensor pairs so that the lower bound on the variance of an unbiased estimator, known as the Cram\'er--Rao bound~(CRB), is minimized.
Since the CRB is proportional to the noise variance~\cite{yaqin2025sensor}, it increases as the target-to-sensor distance becomes larger.
Static pairing~\cite{yaqin2025sensor} selects a set of pairs by minimizing the determinant of the CRB averaged over the region of interest~(ROI), and fixes it during tracking.
Dynamic pairing~\cite{ikura2026misocp} re-selects the pairs at each time to minimize the CRB at the current estimate instead of averaging it over the ROI.

They assume that the position of each sensor is fixed.
This implies that re-selecting pairs alone can improve the CRB slightly when the target is far from most sensors.

\noindent
\textbf{Sensor Relocation}: 
Sensor relocation is formulated as a trajectory optimization of each sensor.
\cite{meng2016optimal} updates the sensor positions at each time.
This problem is formulated as a minimization of the trace of the CRB under sensor movement constraints.
In this setting, the sensor pairs do not change during tracking.

\noindent
\textbf{Summary}: 
In summary, existing designs do not fully address the combined problem of sensor pairing and relocation: The best pairs are variable with the current sensor positions, and the best positions in turn depend on the active pairs.

\section{Simultaneous Pairing and Relocation for Target Tracking}
\label{sec:proposed}
In this section, we present the proposed method that optimizes the active sensor pairs and the relocation simultaneously at each time.

We review the framework of the proposed online target tracking shown in Fig.~\ref{fig:framework}.
Hereafter, we use the subscript $t$ to denote the time.
Our goal is to estimate the target position $\hat{\bm{p}}_t$ from $\bm{z}_{\mathcal{M}_t}(\bm{p})$.
First, sensor pairing and relocation are performed.
The inputs are the previous sensor positions $\bm{S}_{t-1}=[\bm{s}_{1,t-1},\dots,\bm{s}_{N,t-1}]^{\top}$ and the estimated target position at the previous time, $\hat{\bm{p}}_{t-1}$.
The outputs are the active pairs $\mathcal{M}_t$ and the new sensor positions $\bm{S}_t$.
Then, sensors move to their new positions and activate the pairs in $\mathcal{M}_t$ to obtain the TDOAs $\bm{z}_{\mathcal{M}_t}(\bm{p})$.
Finally, it estimates the current target position $\hat{\bm{p}}_t$ based on $\bm{z}_{\mathcal{M}_t}(\bm{p})$.
These three operations repeat at each time.
We assume that the target moves only a limited distance between consecutive time slots.

\subsection{Formulation of Simultaneous Optimization Problem}
At time $t$, we select a set of sensor pairs $\mathcal{M}_t$ and the new sensor positions $\bm{S}_t$ that maximize the localization accuracy using $\mathbf{F}_{\mathcal{M}}(\hat{\bm{p}}_{t-1},\bm{S}_t)$.

Here, an MSN is described by an unweighted graph whose nodes are the sensors and whose edges are the sensor pairs.
Its node set $\mathcal{V}$ contains the $N$ sensors, and its edge set $\mathcal{E}$ holds the $\binom{N}{2}$ candidate pairs, indexed by $k$.
An active set of sensor pairs is a subset $\mathcal{M}_t\subseteq\mathcal{E}$ with $|\mathcal{M}_t|=K$.
The sensors move within the region of interest $\mathcal{A}\subset\mathbb{R}^2$, which we assume to be convex, and we assume all of sensor coordinates are not exactly identical to $\hat{\bm{p}}_{t-1}$ to keep $\bm{w}_k$ well defined.

Since $\Sig_{\mathcal{M}_t}$ is diagonal, $\Sig_{\mathcal{M}_t}^{-1}=\operatorname{diag}\big(\{1/\sigma_k^2\}_{k\in\mathcal{M}_t}\big)$ and $\Sig_{\mathcal{M}_t,a}=\operatorname{diag}\big(\{[\bm{g}_k]_a\}_{k\in\mathcal{M}_t}\big)$, where $\bm{g}_k\triangleq\nabla_{\bm{p}}\sigma_k^2=2\sigma_0^2\big(2\hat{\bm{p}}_{t-1}-\bm{s}_i-\bm{s}_j\big)$.
Substituting these into~\eqref{eq:fimT} gives
\begin{equation}
\label{eq:fimentry}
  [\mathbf{F}_{\mathcal{M}_t}(\hat{\bm{p}}_{t-1},\bm{S}_t)]_{ab}=\sum_{k\in\mathcal{M}_t}\Big(\frac{[\bm{w}_k]_a[\bm{w}_k]_b}{\sigma_k^2}
  +\frac{[\bm{g}_k]_a[\bm{g}_k]_b}{2\sigma_k^4}\Big).
\end{equation}
As a result, the matrix form of~\eqref{eq:fimentry} is given by
\begin{equation}
\label{eq:fimsum}
  \mathbf{F}_{\mathcal{M}_t}(\hat{\bm{p}}_{t-1},\bm{S}_t)=\sum_{k\in\mathcal{M}_t}\mathbf{B}_k(\bm{S}_t),~~
  \mathbf{B}_k(\bm{S}_t)=\frac{\bm{w}_k\bm{w}_k^\top}{\sigma_k^2}
  +\frac{\bm{g}_k\bm{g}_k^\top}{2\sigma_k^4}.
\end{equation}

As in~\cite{yaqin2025sensor}, we measure the localization information by the D-optimality criterion $\det(\mathbf{F}_{\mathcal{M}_t})$.
Therefore, the problem solved at time $t$ is formulated as:
\begin{subequations}
\label{eq:joint}
\begin{align}
 (\mathcal{M}_t,&\bm{S}_t)=\underset{\mathcal{M},\,\bm{S}}{\operatorname{argmin}}~
   f(\mathcal{M},\bm{S})\triangleq
   -\det\!\big(\mathbf{F}_{\mathcal{M}}(\hat{\bm{p}}_{t-1},\bm{S})\big)
   \label{eq:joint_obj}\\
 \text{s.t.}~~
    &|\mathcal{M}|=K,~~\bm{d}(\mathcal{M})\le\Dmax\bm{1}, \label{eq:c_pair}\\
    &\|\bm{s}_i-\bm{s}_{i,t-1}\|\le\dmove,~\bm{s}_i\in\mathcal{A},~
     \|\hat{\bm{p}}_{t-1}-\bm{s}_i\|\ge\rmin,~\forall i, \label{eq:c_place}
\end{align}
\end{subequations}
where $[\bm{d}(\mathcal{M})]_i$ is the degree of sensor $i$ in $(\mathcal{V},\mathcal{M})$.
The degree budget $\bm{d}(\mathcal{M})\le\Dmax\bm{1}$ caps the number of communication links~\cite{yaqin2025sensor}.
The mobility budget $\|\bm{s}_i-\bm{s}_{i,t-1}\|\le\dmove$ limits how far a sensor can travel in one frame.
The region constraint $\bm{s}_i\in\mathcal{A}$ confines every sensor to the region of interest.
The keep-out constraint $\|\hat{\bm{p}}_{t-1}-\bm{s}_i\|\ge\rmin$ holds every sensor at least the keep-out radius $\rmin$ away from the target estimate.

\subsection{Solver}
\label{subsec:solver}
The problem~\eqref{eq:joint} is non-convex in both $\mathcal{M}$ and $\bm{S}$ and thus it is difficult to solve directly.
Instead, we alternately solve two subproblems: we repeatedly update the pairing $\mathcal{M}$ with the relocation $\bm{S}$ fixed and then update $\bm{S}$ with $\mathcal{M}$ fixed.
We show each subproblem in detail below.

\noindent\textit{Pairing subproblem (fixed $\bm{S}$).}~ This subproblem chooses the $K$ pairs that minimize $f(\mathcal{M},\bm{S})$, under the cardinality and degree constraints of~\eqref{eq:c_pair}.
We reformulate the selection problem as a MISOCP that a branch-and-bound algorithm solves to a global optimum~\cite{sagnol2015misocp}\footnote{For more details, we refer to our companion paper~\cite{ikura2026misocp}.}.

\noindent\textit{Relocation subproblem (fixed $\mathcal{M}$).}~ The goal of this subproblem is to move the sensors so as to minimize $f(\mathcal{M},\bm{S})$ under the constraints~\eqref{eq:c_place}.
Since $f$ is non-convex in $\bm{S}$, it is intractable to directly minimize $f$.
We instead apply an iterative solver based on majorization-minimization~\cite{levitin1966constrained}.

We compute quadratic upper-bound of $f$ around the current iterate $\bm{S}^{(q)}$ with a sufficiently large parameter $\rho>0$~\cite{sun2017majorization}:
\begin{equation}
\label{eq:major}
  f(\bm{S})\le f(\bm{S}^{(q)})+\operatorname{tr}\big(\nabla f(\bm{S}^{(q)})^\top(\bm{S}-\bm{S}^{(q)})\big)+\tfrac{\rho}{2}\|\bm{S}-\bm{S}^{(q)}\|_F^2.
\end{equation}
One iteration of the algorithm is obtained by minimizing the right-hand side of~\eqref{eq:major} over the feasible set.
It is computed by the following two steps.

1.  We compute the gradient descent with step size $1/\rho$:
\begin{equation} \label{eq:placement-update}
  \bm{s}_i^{\star}=\bm{s}_i^{(q)}-\nabla_{\bm{s}_i}f/\rho.
\end{equation}
Using the chain rule, the gradient of $f$ with respect to $\bm{s}_i$ is:
\begin{equation}
\label{eq:partials}
  [\nabla_{\bm{s}_i}f]_\ell
  =-\det(\Fmat)\!\!\sum_{k\in\mathcal{E}_i\cap\mathcal{M}}\!\!
    \operatorname{tr}\!\big(\Fmat^{-1}\partial_\ell\mathbf{B}_k\big),  ~~\ell\in\{1,2\},
\end{equation}
where
\begin{equation}
\begin{split}
  \partial_\ell\mathbf{B}_k
  =&\frac{\partial_\ell\bm{w}_k\bm{w}_k^\top+\bm{w}_k\partial_\ell\bm{w}_k^\top}{\sigma_k^2}
  -\frac{\bm{w}_k\bm{w}_k^\top\,\partial_\ell\sigma_k^2}{\sigma_k^4}\\
  &+\frac{\partial_\ell\bm{g}_k\bm{g}_k^\top+\bm{g}_k\partial_\ell\bm{g}_k^\top}{2\sigma_k^4}
  -\frac{\bm{g}_k\bm{g}_k^\top\,\partial_\ell\sigma_k^2}{\sigma_k^6},\\
\end{split}
\end{equation}
in which $\partial_\ell\bm{w}_k=-\frac{(\mathbf{I}-\bm{u}_i\bm{u}_i^\top)\bm{e}_\ell}{|\hat{\bm{p}}_{t-1}-\bm{s}_i|},~\partial_\ell\bm{g}_k=-2\sigma_0^2\bm{e}_\ell,$ and $
  \partial_\ell\sigma_k^2=-2\sigma_0^2 |\hat{\bm{p}}_{t-1}-\bm{s}_i|\,\bm{u}_i^\top\bm{e}_\ell$.
Here, $\mathcal{E}_i \subset \mathcal{E}$ is the set of pairs which contain sensor $i$, $\bm{e}_1=[1,0]^\top$ and $\bm{e}_2=[0,1]^\top$, and $\partial_\ell = \partial/\partial[\bm{s}_i]_\ell$.

2. We compute the projection onto the feasible set:
\begin{equation}
    \bm{s}_i^{(q+1)}=\Pi_{\mathcal{C}_i}\big(\bm{s}_i^{\star}\big),~\forall i,
\end{equation}
where
\begin{equation}
\label{eq:cset}
  \mathcal{C}_i=(\mathcal{B}_i\cap\mathcal{A})\setminus\mathcal{D},
\end{equation}
in which $\mathcal{B}_i=\{\bm{x}:\|\bm{x}-\bm{s}_{i,t-1}\|\le\dmove\},$ and $
  \mathcal{D}=\{\bm{x}:\|\bm{x}-\hat{\bm{p}}_{t-1}\|<\rmin\}$.
The set $\mathcal{C}_i$ in~\eqref{eq:cset} removes the keep-out disk $\mathcal{D}$ from $\mathcal{B}_i\cap\mathcal{A}$.

Generally, $\mathcal{C}_i$ is a non-convex set. Nevertheless, the projection $\Pi_{\mathcal{C}_i}$ in~\eqref{eq:placement-update} is efficiently computed as follows.
We first project $\bm{s}_i^{\star}$ onto the convex part $\mathcal{B}_i\cap\mathcal{A}$ through Dykstra's algorithm~\cite{boyle1986projections}, which gives $\bm{y}_i=\Pi_{\mathcal{B}_i\cap\mathcal{A}}(\bm{s}_i^{\star})$.
Then, if $\bm{y}_i$ falls inside $\mathcal{D}$, $\bm{s}_i^\star$ is projected onto the nearest point of the boundary $\partial\mathcal{D}$ that also lies in $\mathcal{B}_i\cap\mathcal{A}$.
It is formally written as:
\begin{equation}
\label{eq:proj}
  \Pi_{\mathcal{C}_i}(\bm{s}_i^{\star})=
  \begin{cases}
    \bm{y}_i, & \bm{y}_i\notin\mathcal{D},\\
    \arg\min_{\bm{x}\in\partial\mathcal{D}\cap\mathcal{B}_i\cap\mathcal{A}}\|\bm{x}-\bm{s}_i^{\star}\|, & \bm{y}_i\in\mathcal{D}.
  \end{cases}
\end{equation}
The projection onto $\partial \mathcal{D}$ can be computed as:
\begin{equation}
\label{eq:proj2}
  \underset{\bm{x}\in\partial\mathcal{D}}{\text{argmin}}\|\bm{x}-\bm{s}_i^{\star}\|
  =\hat{\bm{p}}_{t-1}+\rmin\frac{\bm{s}_i^{\star}-\hat{\bm{p}}_{t-1}}{\|\bm{s}_i^{\star}-\hat{\bm{p}}_{t-1}\|}.
\end{equation}
If this point further falls out of $\mathcal{B}_i\cap\mathcal{A}$, the solution is one of the intersections of $\partial\mathcal{D}$ and the boundary of $\mathcal{B}_i\cap\mathcal{A}$.
Since such intersections are finite, we obtain the final solution by comparing the distances from these points to $\bm{s}_i^{\star}$.
We iterate this update~\eqref{eq:placement-update} until the convergence.

Algorithm~\ref{alg:joint} summarizes the proposed simultaneous pairing and relocation at time $t$.
For the parameter $\rho$, we start it at $\max_i\|\nabla_{\bm{s}_i}f\|/\dmove$, and double it until~\eqref{eq:major} is satisfied.
This algorithm is guaranteed to converge to a critical point\footnote{A convergence analysis is omitted due to space limitations and will be given in the journal version of this paper.}.

\begin{figure*}[t]
  \centering
  \includegraphics[width=0.88\textwidth]{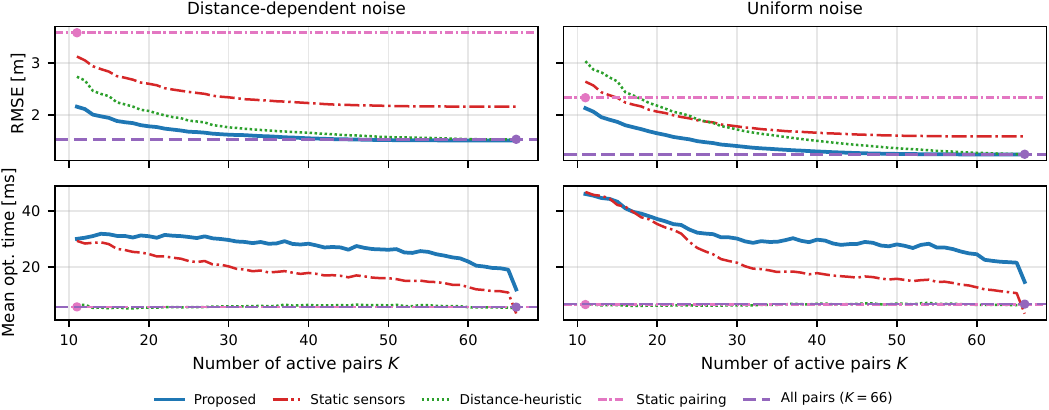}
  \caption{Accuracy (top row, time-averaged RMSE in m) and cost (bottom row, mean per-frame optimization time in ms) against the number of active pairs $K\in\{11,\dots,66\}$, under distance-dependent (left column) and uniform (right column) noise.
  Lower is better in both rows.
  The three curves are swept over $K$.
  The two horizontal lines are the methods whose number of pairs is fixed by their designs; the markers indicate that number ($K=11$ for static pairing, $K=66$ for all pairs).}
  \label{fig:combined}
\end{figure*}

\begin{algorithm}[t]
\caption{Simultaneous sensor pairing and relocation}
\label{alg:joint}
\begin{algorithmic}[1]
\REQUIRE $\hat{\bm{p}}_{t-1}$, $\bm{S}_{t-1}$, $\mathcal{M}_{t-1}$, $K,\Dmax,\dmove,\rmin,\mathcal{A},\epsilon_1,\epsilon_2$.
\STATE Set $\bm{S}^{(p)}\!\leftarrow\!\bm{S}_{t-1}$, $\mathcal{M}^{(p)}\!\leftarrow\!\mathcal{M}_{t-1}$, $p\leftarrow 0$
\WHILE{$|f(\mathcal{M}^{(p)},\bm{S}^{(p)})-f(\mathcal{M}^{(p-1)},\bm{S}^{(p-1)})|\ge\epsilon_1$}
  \STATE Pairing: solve the pairing subproblem at $\bm{S}^{(p)}$, giving $\mathcal{M}^{(p)}$
  \STATE Relocation: set $\bm{S}^{(q)}\leftarrow\bm{S}^{(p)}$, $q=0$
  \WHILE{$|f(\mathcal{M}^{(p)},\bm{S}^{(q)})-f(\mathcal{M}^{(p)},\bm{S}^{(q+1)})|\ge\epsilon_2$}
    \STATE Compute $\nabla f$ by~\eqref{eq:partials} and update $\bm{S}^{(q)}$ by~\eqref{eq:placement-update} and~\eqref{eq:proj}
    \STATE $q\leftarrow q+1$
  \ENDWHILE
  \STATE $\bm{S}^{(p+1)}\leftarrow\bm{S}^{(q)}$
  \STATE $p \leftarrow p+1$
\ENDWHILE
\RETURN $\mathcal{M}_t\!\leftarrow\!\mathcal{M}^{(p)}$, $\bm{S}_t\!\leftarrow\!\bm{S}^{(p)}$
\end{algorithmic}
\end{algorithm}

\section{Experiments}
\label{sec:experiments}
In this section, we perform a target tracking simulation with a synthetic mobile sensor network.

\subsection{Setting}
In a square region $\mathcal{A}=[0,1000]\times[0,1000]$\,m, $N=12$ sensors are initially placed at random with a minimum separation of $30$\,m, and one target moves inside $\mathcal{A}$.
The target follows a random waypoint motion~\cite{bettstetter2004stochastic}: It moves straight toward a destination drawn $30$ to $150$\,m away in a uniformly random direction at a speed drawn uniformly from $8$ to $32$\,m/s, pauses for up to $1$\,s on arrival, and then draws the next destination.
Its heading and speed therefore change at every waypoint.
The control frame is $\Delta t=0.25$\,s~\cite{watson1994tracking}, and the target is tracked for $100$ frames.
A sensor moves at most $\dmove=2$\,m per frame.
This is a quarter of the maximum distance that the target moves in one frame.
The radius of the keep-out disk is set as $\rmin=\dmove/\sqrt{2}=\sqrt{2}$\,m, which is the largest radius for which~\eqref{eq:c_place} remains feasible at each time.

We sweep $K$ from $11$ to $66$ in unit steps with the degree budget $\Dmax(K)=\min(\lceil K/3\rceil, N-1)$.
Following~\cite{yaqin2025sensor}, we take $K=N-1=11$ as the lower limit.
We compare two noise regimes with all the other settings identical.
The first is the distance-dependent model~\eqref{eq:noise} with $\sigma_{\min}=3$\,m and $\sigma_0=10^{-2}$.
The second is a uniform $\sigma=6$\,m ($\sigma_0=0$).
The convergence thresholds are set to $\epsilon_1=\epsilon_2=10^{-4}$.
The target position is estimated using the Gauss--Newton method~\cite{foy1976position,torrieri2007statistical}.

Two of the baselines are swept over $K$ under the same estimator and mobility constraints as the proposed method.
\begin{itemize}
  \item \textbf{Static sensors~\cite{ikura2026misocp}:} This baseline runs the same MISOCP pairing solver as the proposed method without sensor relocation.
    This is the design of our companion work.
  \item \textbf{Distance-heuristic:} It selects the pairs of smallest target-to-pair distance under the same degree budget.
  Its relocation is obtained in the same way as the proposed method.
\end{itemize}
The remaining two methods use the number of pairs fixed by their original designs.
\begin{itemize}
  \item \textbf{Static pairing~\cite{yaqin2025sensor,meng2016optimal}:} The pair set is chosen offline by greedy maximization of a spatially averaged D-optimality criterion under per-node resource limits~\cite{yaqin2025sensor}. This design uses only $N-1=11$ pairs. Then the sensors are relocated online following~\cite{meng2016optimal} as described in Sec.~\ref{sec:related}.
  \item \textbf{All pairs:} All $66$ pairs are active, and the sensor relocation is obtained in the same way as the proposed method. This baseline serves as an unconstrained reference.
\end{itemize}

For each method, we also report the mean per-frame optimization time.
All codes are written in Python, and timings are measured on an Apple MacBook Pro with an M2 Pro chip.
We test $200$ independent runs and average the results.

\begin{figure}[t]
  \centering
  \includegraphics[width=0.98\columnwidth]{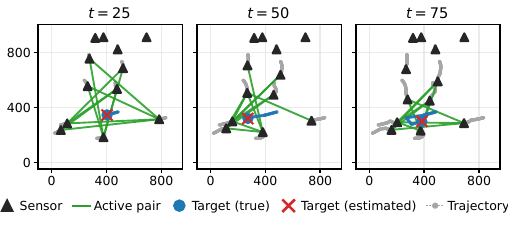}
  \caption{Snapshots of one trial of the proposed method at $t=25$, $50$, and $75$ ($K=11$, distance-dependent noise).
  Each panel shows the sensors (black triangles), the active pairs (green lines), the true target (blue dot), its estimate (red cross), and the trajectories (dotted).}
  \label{fig:steps}
\end{figure}

\subsection{Results}
The top row of Fig.~\ref{fig:combined} shows the time-averaged RMSEs.
We observe that the proposed method shows better performance than all other methods except for All pairs.
The proposed method is effective especially when $K$ is small, i.e., when the communication budget is tight.
The proposed curve also approaches All pairs as $K$ grows and reaches it about $K\approx40$--$50$.

The bottom row of Fig.~\ref{fig:combined} reports the mean per-frame optimization time.
We observe that the mean execution time of the proposed method is larger than those of the baselines while it is still significantly shorter than $\Delta t$.

Fig.~\ref{fig:steps} shows that the proposed method activates the pairs of the sensors near the target.
Each sensor moves toward the target only if it is involved in an active pair.
The sensors far from the target's path are not activated during the trial.

\section{Conclusion}
\label{sec:conclusion}
We proposed a simultaneous optimization of sensor pairing and relocation for TDOA-based online target tracking.
Both the sensor pairing and the relocation problems are cast as one problem under a D-optimality criterion on the FIM.
We solve it by alternating the MISOCP for pairing with a majorization-minimization for relocation.
Simulation results demonstrate that the proposed method obtains improved tracking accuracy compared with alternative methods, within a practical time for online tracking.

\clearpage
\bibliographystyle{IEEEbib_initials}
\bibliography{TDOA}

\end{document}